\documentclass[preprint,aps,amssymb,showpacs]{revtex4}
\begin{document}
\preprint{IMSc/2002/10/35}

\title{A Discrete Time Presentation of Quantum Dynamics}

\author{G. Date}
\email{shyam@imsc.res.in}
\affiliation{The Institute of Mathematical Sciences\\
CIT Campus, Chennai-600 113, INDIA.}
\begin{abstract}

Inspired by the discrete evolution implied by the recent work on loop
quantum cosmology, we obtain a discrete time description of usual
quantum mechanics viewing it as a constrained system. This description,
obtained
without any approximation or explicit discretization, 
mimics features of the discrete time evolution of loop quantum cosmology. 
We discuss the continuum limit, physical inner product and matrix elements 
of physical observables to bring out various issues regarding viability of 
a discrete evolution. We also point out how a continuous time could
emerge without appealing to any continuum limit.

\end{abstract}


\maketitle

\section{Introduction}
The recent work of Bojowald on Loop Quantum Cosmology (LQC) obtains
the Wheeler-DeWitt equation (Hamiltonian constraint) as a difference
equation \cite{wd-oprs}. The eigenvalues of the volume operator are 
discrete in quantum geometry and are taken as playing the role of a 
`discrete time' in the context of isotropic LQC \cite{discrete}. The 
order of the difference equation is typically high (16 for isotropic, 
Bianchi-I) and the number of independent (and non-degenerate) solutions is 
reduced by one due 
to the coefficient of the highest (lowest) order term vanishing for 
discrete time equal to zero \cite{isotropic}. This feature is crucial for 
the `singularity avoidance' mechanism \cite{absense}. Furthermore, a 
continuum limit is defined wherein the Immirzi parameter plays a crucial 
role. This limit is used to distinguish the so called `pre-classical' 
solutions and it is shown that the singularity avoidance mechanism also 
leads to a unique (up to normalization) pre-classical solution of the 
Wheeler-DeWitt equation \cite{unique,semiclassical}. These are very 
interesting indications that the procedures of loop quantum gravity as 
applied to cosmological mini-superspaces, do lead to physically reasonable 
solutions. Although certain choices of definitions of constraint operators 
with certain factor orderings can be made with reasonable justification, the
procedure is not devoid of ambiguities \cite{isotropic,ambiguities}. While 
there are many issues to be resolved, we focus on one feature, namely 
`discrete evolution', which seems to be robust. \\

For instance, mathematically, solutions of the Wheeler-DeWitt equation
can be presented as a
sequence of states (isotropic LQC). The sequence label is highly
suggestive of a discrete `time'. This dynamical {\it interpretation} is
something additionally attempted and its viability needs to be established. In
a quantum theory, an evolution interpretation must be established at
least at the level of expectation values, the states being not directly
observable. Furthermore it is not enough to generate a family,
continuous or discrete, of expectation values. It should be possible to
detect the changes paying due attention to the uncertainties. By
contrast, a continuum limit does not appear to be essential for a dynamical
interpretation even though emergence of a continuous time description in a
semi-classical limit is of course desirable. Presently, in the context
of LQC, these issues are discussed somewhat schematically. One of the
motivations for the present work is to have simple, well known examples which 
nonetheless mimic steps taken in LQC. \\

It is well known that the usual quantum mechanical systems (or classical
for that matter) can also be viewed as a constrained system leading to
a frozen-time description. The Dirac procedure nonetheless allows one to
{\it interpret} the physical state condition as an evolution equation - 
the usual Schrodinger equation. By using the `Fock' representation
instead of the Schrodinger representation for the extra time degree of
freedom, one can get a discrete `evolution' equation which mimics all
the features seen in the LQC work. The order of the difference equation
is two but there is a reduction of number of non-trivial independent
solutions; there
is a parameter analogous to the Immirzi parameter which can play a similar
role in exploring pre-classicality and a continuum limit. \\

To explore evolution at the level of expectation values of course needs 
definitions of physical inner products, observables and their matrix elements.
The advantage in the quantum mechanical case however is that we know physical 
inner products and physical matrix elements so that we can push further the 
interpretation of the difference equation as ``really" an evolution equation. 
Furthermore since the Schrodinger and the `Fock' representations are 
equivalent, we can relate the continuous and the discrete evolutions by a 
transform. It turns out that the dynamical interpretation is not as
straightforward as indicated by the LQC works. \\

`Discrete Time' has appeared in the literature several times \cite{history} 
in various forms and with various motivations. The present work is
very different from these earlier works. In particular, we are {\it not
seeking} a discrete time formulation, ab initio, for any particular reason. We
observe that in a frozen time formulation of dynamics, a `time' appears
as a basis label which has an arbitrariness about it. The dynamics is
then obtained as a family of states labeled by the `time'. One has
natural choices of continuous and discrete labels. The choice of the
continuous label leads to the usual quantum dynamics and we explore the
discrete choice in detail, in particular with regards to observability
of evolution. \\

In section II we detail the case of usual quantum mechanics cast in a
frozen time form, both with continuous and discrete time and exhibit
its analogy with LQC. We discuss the continuum limit and show the relation
between the continuous and the discrete time
descriptions. In section III we discuss natural candidates for physical
quantities needed to push the evolution interpretation further and the 
difficulties encountered. In the last section we discuss the issue of 
interpretation in some generality and point out a possible
role of the parameter $\beta$ appearing in the discrete
description. We conclude by making a series of remarks. 

\section{Non-relativistic Quantum Mechanics}

\subsection{Frozen Time Description: Continuum Case}

Let $\Gamma_0$ denote a classical phase space and let ${\cal{H}}_0$ denote
the Hilbert space of the corresponding quantum system. Let $\tau,
\pi_{\tau}$ denote two extra phase space variables corresponding to the
usual continuous time and its `conjugate momentum'. Let $\Gamma_{kin} :=
R^2 \times \Gamma_0 $ denote an extended phase space and let
${\cal{H}}_{kin} := L_2(R,d\tau) \otimes {\cal{H}}_0$ the corresponding
kinematical Hilbert space. At the classical level we impose the single
constraint $\phi := \pi_{\tau} + H(\omega) \approx 0$. Here, $\omega$ denote the
usual phase space coordinates and $H(\omega)$ denotes a time independent
Hamiltonian on $\Gamma_0$. Quantum mechanically, the operator version of
the constraint is imposed on to select the `physical states'. Explicitly, 
the physical states are those on which the operator $\pi_{\tau} + H$ vanishes. 
The constraint operator is of course is {\it not} identically zero.
For
definiteness one may think of $\Gamma_0 = R^{2N},  {\cal H}_0 =
L^2(q^i, d^nq)$ and $H(\omega) = \frac{p_i^2}{2m} + V(q^i)$ though 
this is not necessary. There is no external time any more. \\

At the classical level, the Dirac observables, $A(\tau, \pi_{\tau},
\omega)$ are defined by $\{A, \phi\}_{PB} \approx 0$. Some simple examples
of Dirac observables are: functions of only $\pi_{\tau}$ and functions
independent of $\tau, \pi_{\tau}$ which Poisson commute with the
Hamiltonian (in particular the Hamiltonian itself). This is a rather
limited class of observables. One could however {\it choose} a $\tau$
dependent family of functions of $\omega$ satisfying the differential 
equation $\frac{\partial A}{\partial \tau} + \{A(\tau, \omega), 
H(\omega)\}_{PB} = 0$ with the initial condition $A(0, \omega) =
A_0(\omega)$. Such solutions of the differential equation are trivially
Dirac observables. In particular, the usual solutions of Hamilton's
equation are also Dirac observables but with $\tau \rightarrow - \tau$.
These are the `evolving' observables in a frozen time formulation
\cite{evolving}. \\

In the quantum description, one quantizes the $\tau, \pi_{\tau}$ in the
usual manner and the constraint equation becomes just the Schrodinger
equation. Explicitly, one can write a general vector in
${\cal{H}}_{kin}$ in the form:
\begin{equation}
|\Psi\rangle = \int d\tau |\tau\rangle \otimes |\Phi(\tau)\rangle
\end{equation}
The kinematical inner product is then given by,
\begin{equation}
\langle\Psi^{\prime}|\Psi\rangle_{kin} = \int d\tau \langle\Phi^{\prime}
(\tau)|\Phi(\tau)\rangle_{0}
\end{equation}
The suffix $0$ refers to the inner product in ${\cal{H}}_0$.\\

Thus physical states are those $|\Psi\rangle$ whose $|\Phi(\tau)\rangle$ 
satisfy the usual time dependent Schrodinger equation.
Introducing $U(\tau) := e^{-\frac{i}{\hbar}\tau \hat{H}}$, we denote 
the solutions of the Schrodinger equation as $|\Phi(\tau)\rangle = 
U(\tau)|\Phi(0)\rangle$. The corresponding $|\Psi\rangle$'s are 
{\it not} normalizable with respect to the kinematical inner product since 
the integrand is independent of $\tau$ rendering $\tau$ integration 
divergent. \\
 
Dirac observables are usually defined as those observables which commute
with the constraints. It may be sufficient to require that
the physical observables commute only weakly with the constraints. In this
case the factor ordering must ensure that the constraint operators act
first i.e. are to the right. Weak commutation then amounts to physical
observables acting invariantly on the physical states.
As an example one can define `evolving observables' as
follows \cite{discrete}. Corresponding to a usual operator, $\hat{O}$ on 
${\cal H}_0$, define a family of operators on the space of 
physical states by,
\begin{equation}
[\hat{O}(\tau)|\Phi\rangle](\tau^{\prime}) :=
U(\tau^{\prime})U(-\tau)\hat{O}|\Phi\rangle(\tau)
\end{equation}

These `evolving observables' are physical in the sense they act invariantly 
on the space of physical states. In general, they do {\it not} commute with 
the constraint operator in the full ${\cal H}_{kin}$. Since physical states 
are not kinematically normalizable, one has to define a physical inner product 
on the space of solutions of the constraint. A natural definition suggests 
itself as:
\begin{equation}
\langle \Psi^{\prime}|\Psi\rangle_{phy} := 
\langle \Phi^{\prime}(\tau_0) | \Phi(\tau_0) \rangle_0 %= 
%\int_{R^N}{\phi^{\prime}}^*(\tau_0, q^i) \phi(\tau_0, q^i) 
~~~~~~~{\mbox{$\tau_0$ is some fixed time}}
\end{equation}

The kinematical inner product is just the integral of the physical inner
product over $\tau_0$. It is obvious that the physical inner product is
independent of the particular $\tau_0$ chosen and of course coincides
with the usual inner product in ${\cal H}_0$.\\

It is easy to see that the physical matrix elements of these evolving
observables between physical states are given by,
\begin{equation}
\langle \Psi^{\prime} | \hat{O}(\tau) | \Psi \rangle_{phy} = 
\langle \Phi^{\prime}(0) | U(-\tau) \hat{O} U(\tau) | \Phi(0) \rangle_0
\end{equation}

Hence, the physical matrix elements are obtained as the usual matrix
elements of operators on ${\cal{H}}_0$. We see thus that the usual 
description of quantum dynamics in terms of a continuous time can be 
recast as a frozen time presentation. This is of course well known. We 
will now introduce a discrete time description which mimics all the 
features seen in the loop quantum cosmology.

\subsection{Discrete case}

Instead of choosing the usual Schrodinger representation for $\tau,
\pi_{\tau} = - i \hbar \frac{\partial}{\partial \tau}$, let us introduce a 
number representation. Define $a := \alpha \tau + i \beta \pi_{\tau}$
and $a^{\dagger}$ its hermitian conjugate. $\alpha, \beta$ are real and
satisfy $2 \alpha \beta \hbar = 1$. Hence we have a one parameter family
of creation-annihilation operators labeled by $\beta$, say. This
parameter is expected to play a role analogous to that played by the Immirzi 
parameter in LQC. We will choose the eigenvalues of the number operator, 
$N := a^{\dagger} a$ as our discrete time label.  Notice that these eigenvalues 
are independent of $\beta$. For future use in continuum limit we note that,
\begin{equation} 
N ~ = ~ (\frac{1}{2 \hbar \beta})^2 \tau^2 + \beta^2 \pi^2_{\tau} - 
\frac{1}{2} ~~~~~ \sim ~~~~~ \frac{1}{4 \hbar^2 \beta^2} 
\tau^2 ~~~~~~~ \mbox{as $\beta \rightarrow 0$}
\end{equation} 

This will justify the eigenvalues of the number operator being
identified with $\tau$ at least for large eigenvalues and small $\beta$
and of course $n$ is monotonic in $\tau^2$. \\

In terms of creation-annihilation operators, the constraint equation
becomes,
\begin{equation}
\left(\frac{a - a^{\dagger}}{2 i \beta} + \hat{H}\right) | \Psi \rangle = 0 .
\end{equation}

Writing $|\Psi\rangle = \sum_{n = 0}^{\infty} |n\rangle \otimes
|\Phi_n\rangle $ one obtains the kinematical inner product as,
\begin{equation}
\langle\Psi^{\prime}|\Psi\rangle 
= \sum_{n = 0}^\infty \langle \Phi^{\prime}_n | \Phi_n \rangle_0.
\end{equation}

The physical states are then those $|\Psi\rangle$ whose $|\Phi_n\rangle$'s
satisfy,
\begin{equation}
|\Phi_{n + 2}\rangle  =  \frac{-2 i \beta}{\sqrt{n + 2}} \hat{H} |\Phi_{n +
1}\rangle + \sqrt{\frac{n + 1}{n + 2}} | \Phi_n \rangle ~~~~ \forall ~~n
\ge -1 
\end{equation}

This is our discrete time evolution (discrete Schrodinger equation).
We notice that the difference equation is an operator difference
equation of order two implying that two vectors (in ${\cal{H}}_0$ ) have
to be specified to determine a solution i.e. has `two independent' solutions. 
However exactly as in the case of loop quantum
cosmology, we have a consistency condition (since spectrum of $N$ is
bounded below) which fixes $|\Phi_1\rangle$ in terms of $|\Phi_0\rangle$
implying a unique solution for every given $|\Phi_0\rangle$. By contrast, 
in the continuum description, there is no such condition but the
equation is of course a first order differential equation. \\

For subsequent analysis, it is convenient to convert the vector equation 
in to infinitely many scalar equations by expanding the $|\Phi_n\rangle = 
\sum_{\alpha} C_n^{\alpha}|{\cal E}_{\alpha}\rangle$ in the eigenbasis 
of the $\hat{H}$. For simplicity we have assumed that spectrum of $\hat{H}$ 
in ${\cal H}_0$ is discrete. The discrete equation then becomes,
\begin{equation}
C^{\alpha}_{n + 2} = - \frac{2 i \beta {\cal{E}_{\alpha}}}{\sqrt{n + 2}} 
C^{\alpha}_{n + 1}  + \sqrt{\frac{n + 1}{n + 2}} C^{\alpha}_n
~~~~~~~~~ \forall ~ n \ge -1, ~~ \forall ~ \alpha.
\end{equation}

Let us now turn to the ``continuum limit". This can be understood in
various ways. A simple way is to ask whether one can find continuous
variable $t$  and a function $C^{\alpha}(t)$ which will interpolate a
solution of the discrete equation  for large $n$. To explore this 
let us look for a function $C(t)$  and a function $t(n)$ 
such that in a suitable large $n$ limit one has (suppressing the label
$\alpha$),
\begin{equation}
C_{n + k} := C(t(n + k)) ~:=~ C(t(n) + k \delta t) ~~ \approx ~~ C(t) + k
\delta t \frac{\partial C}{\partial t}.
\end{equation}

where, $k \delta t = t(n + k) - t(n)$ has been used.  Treating
$\delta t$ small is equivalent to requiring $t(n)$ to be slowly varying
with $n$. Substitution in the difference equation and keeping terms to leading 
order in $n$ (treating $\delta t$ also as small), leads to 
$C^{-1}\frac{\partial C}{\partial t} ~ = ~ -i {\cal{E}}\frac{\beta}{\delta t 
\sqrt{n}}$.
The left hand side is a function of $t$ by assumption so we must have
$\frac{\beta}{\delta t \sqrt{n}}$ to be a function of $t$. Since the
limit is to be considered for all $C^{\alpha}$, we have excluded ${\cal{E}}$. 
Simplicity and dimensional considerations then suggest that
we choose $t$ such that $\frac{\beta}{\delta t \sqrt{n}} := \hbar^{-1}$.
The variable $t$ so specified will be denoted by $\tau$. It follows that
$\tau(n) = 2 \hbar \beta \sqrt{n}$.  Observe that $\delta\tau \sim
n^{-1/2}$  and thus vanishes for large $n$ {\it without} having to take
$\hbar\beta$ to be vanishingly small. This is different from the LQC.
To get a finite $\tau$ for arbitrarily large $n$ however, we must consider a
joint limit $n \rightarrow \infty , \beta \rightarrow 0$ keeping $\tau$
fixed. We could have taken $\hbar\beta$ to zero but presently we are
interested in a continuum limit instead of a semi-classical limit. It
follows that in the above joint limit $C(\tau)$ satisfies the usual
Schrodinger equation and so does the full $|\Phi\rangle(\tau)$.  Thus we
see that continuous functions that interpolate solutions of the
difference equation asymptotically are solutions of the usual Schrodinger equation with
a suitable identification of $\tau(n)$.\\

Can we obtain $C(\tau)$ as a limiting function from the joint limit of
$C_n$? The answer is yes. Consider the difference
equation for large $n$ but with fixed $\beta$,
\begin{equation}
C^{\alpha}_{n + 2} - C^{\alpha}_{n}  = - (\frac{2 i
\beta{\cal E}_{\alpha}}{\sqrt{n}}) C^{\alpha}_{n + 1}  + o(\frac{1}{n}).
\end{equation}

The left hand side of this equation equated to zero is an equation with
constant coefficients (Poincare type)\cite{diff-eqns}. The asymptotic behaviour of its 
solutions is given in terms of its characteristic roots obtained by
substituting $C^{\alpha}_n \sim \lambda^n$. The characteristic roots are just 
$\lambda = \pm 1$, independent of the label $\alpha$.  Evidently, the root 
$\lambda = -1$ can not correspond to a solution which has limiting value 
in the joint limit. Furthermore, even for $\lambda = 1$, one can not see
the limit to be a solution of the continuum Schrodinger equation. One
needs a more refined ansatz:  $C_n \sim \lambda^n \mu^{\sqrt{n}}$.
Substitution determines $\lambda = \pm 1$ from the $n$ independent term
and also gives $\ell n \mu = -2 i \lambda\beta {\cal{E}}_{\alpha}$ from
the sub-leading $n^{-1/2}$ term. For $\lambda = 1$ we see that
$C^{\alpha}_n$ goes over to the solution of the continuum Schrodinger
equation. By contrast , $\lambda = -1$ does not have a limit. A generic
asymptotic solution will be a linear combination of these two asymptotic 
solutions
and it will not have a limiting value in the joint limit. There is then a
unique solution that does have limiting value. This is very similar to the
arguments \cite{unique}.\\

So far the steps are completely analogous to those taken in loop quantum
cosmology. However here we run in to a problem. {\it No exact solution of 
the difference equation can possibly have a non-zero and finite limiting value 
in the joint limit.} This follows because the ratio of the $C^{\alpha}_n$ 
for $n$ odd and $n$ even is necessarily purely imaginary as is evident from 
the difference equation. Thus although the assumptions we made about a
conceivable continuum limit do admit a corresponding ansatz for the asymptotic
solution of the difference equation, no exact solution can in fact support 
such an ansatz. The notion of pre-classical limit as articulated in LQC 
\cite{unique,semiclassical}, is not realized by any exact solution even
though a continuum limit in the sense of difference equation going over
to a differential equation is valid. 
One has only {\it asymptotic solutions} i.e. solutions of asymptotic
equation as distinct from asymptotic form of solutions of the exact equation,
which have a pre-classical limit but no exact solution has this
property. It seems that the continuum solutions
can at best be thought of as {\it approximating} the exact solution that
too only asymptotically i.e. for large $n$.\\
%
%One has asymptotic solutions which
%have a pre-classical limit but no exact solution has this property.
%It seems that the continuum solutions can at best be thought of as {\it 
%approximating} the exact discrete solutions that too only asymptotically. \\

Is not having any pre-classical solutions a disaster for getting a
continuum picture? Not necessarily. For emergence of a continuum
description from an underlying discrete one what is needed is a {\it 
mapping} to continuous description and not necessarily a continuum
{\it limit}. The next sub-section shows how this can happen.

\subsection{Relating the continuum and the discrete descriptions}

The evolution equations have been derived by writing:
\begin{eqnarray}
|\Psi\rangle \in {\cal{H}}_{kin} & = & \int_{\infty}^{\infty} d\tau 
|\tau\rangle \otimes |\Phi(\tau)\rangle	\mbox{\hspace{2.0cm} continuous case}
\nonumber \\
& = & \sum_{n = 0}^{\infty} |n\rangle \otimes |\Phi_n\rangle 
\mbox{\hspace{2.9cm} discrete case}
\end{eqnarray}

Imposing the constraint in the Schrodinger representation and the Fock
representation respectively leads to the usual Schrodinger equation for
$|\Phi(\tau)\rangle$ and the difference equation for $|\Phi_n\rangle$.
The two basis vectors $|\tau\rangle$ and $|n\rangle$ are related as:
\begin{eqnarray}
|\tau\rangle & = &  \sum_{n = 0}^{\infty} |n\rangle \langle
n|\tau\rangle  =  \sum_n
f_n^*(\tau) |n\rangle , \nonumber \\
|n\rangle & = & \int_{- \infty}^{\infty} d \tau |\tau\rangle
\langle\tau|n\rangle  =  \int d \tau f_n(\tau) |\tau\rangle
\end{eqnarray}

The transformation functions, $f_n(\tau)$ are easily determined and are
given by,
\begin{eqnarray}
f_n(\tau) & := & \langle\tau|n\rangle = {\cal{N}}_n H_n(\xi) e^{-
\frac{1}{2}\xi^2} \mbox{\hspace{2.0cm} $
\xi ~ := ~ \frac{\tau}{\sqrt{2} ~ \hbar \beta}$,} \nonumber \\
{\cal{N}}_n & = & \left( (2 \pi)^{1/4} \sqrt{\hbar\beta} \sqrt{2^n
n!} \right)^{-1} 
\end{eqnarray}

and $H_n(\xi)$ are the Hermite polynomials. The vectors
$|\Phi(\tau)\rangle$ and $|\Phi_n\rangle$ are easily seen to be related
as:

\begin{eqnarray}
|\Phi(\tau)\rangle & = & \sum_n f_n(\tau) |\Phi_n\rangle \nonumber \\
|\Phi_n\rangle & = & \int d\tau f_n^*(\tau) |\Phi(\tau)\rangle .
\end{eqnarray}

Using the properties of the Hermite polynomials, it is easy to see that
{\it $|\Phi_n\rangle$ satisfy the difference equation iff $|\Phi(\tau)\rangle$
satisfies the Schrodinger equation.} The respective initial states are
related as:
\begin{equation}
|\Phi_0\rangle = \frac{\sqrt{2}}{(2 \pi)^{1/4}} \sqrt{\hbar\beta} 
\left\{ \int_{-\infty}^{\infty} d \xi e^{- \frac{\xi^2}{2} - i \sqrt{2} 
\xi \beta \hat{H}}\right\} |\Phi(0)\rangle .
\end{equation}

By expanding $|\Phi(0)\rangle$ in eigenstates of the Hamiltonian one can
show that,
\begin{equation}
\langle\Phi_0 | \Phi_0\rangle_{_0} \le 2 \sqrt{2 \pi} \hbar\beta
\langle\Phi(0) | \Phi(0)\rangle_{_0} .
\end{equation}

This shows that in the limit $\beta \rightarrow 0$ the transform breaks
down.\\

This route, available in the present case, shows that it is possible to 
define states depending on a continuous variable in a differentiable
manner {\it without} appealing to any `pre-classical' or otherwise
limiting procedure.

\section{Physical Quantities}

As a first attempt, we will just mimic the steps followed in the
continuous time case. For this it is convenient to write the second order 
difference equation as a first order matrix difference equation. This is 
easily achieved \cite{discrete}. The evolution equation can be written as:
\begin{eqnarray}
\left( \begin{array}{c} |\Phi_{n + 1}\rangle \\ |\Phi_{n +
2}\rangle \end{array} \right) & = & \left( \begin{array}{cc} 0 & {\mathbb I}
\\ \sqrt{\frac{n + 1}{n + 2}}{\mathbb I} & \frac{-2 i \beta}{\sqrt{n + 2}}
 \hat{H}
\end{array} \right) \left( \begin{array}{c} |\Phi_{n}\rangle \\ |\Phi_{n +
1}\rangle \end{array} \right) ~~~ \Longleftrightarrow ~~~ z_{n + 1}
= {\mathbb A}(n) z_n  ~~~ n \ge 0 \nonumber \\
\left( \begin{array}{c} |\Phi_{n - 1}\rangle \\ |\Phi_{n 
}\rangle \end{array} \right) & = & \left( \begin{array}{cc} \frac{2 i
\beta}{\sqrt{n}}\hat{H} & \sqrt{\frac{n + 1}{n}} {\mathbb I}
\\ {\mathbb I} & 0
\end{array} \right) \left( \begin{array}{c} |\Phi_{n}\rangle \\ |\Phi_{n +
1}\rangle \end{array} \right) ~~~~~~ \Longleftrightarrow ~~~ z_{n - 1}
= {\mathbb B}(n) z_n ~~~ n \ge 1. 
\end{eqnarray}

It follows from the definitions that ${\mathbb A}(n -1) {\mathbb B}(n) = 
{\mathbb B}(n + 1) {\mathbb A}(n) = {\mathbb I}$. 
Not all states evolving by the above equations are physical though
because the physical states also have to satisfy the consistency condition
namely $|\Phi_1\rangle = -2 i \beta \hat{H} |\Phi_0\rangle$. \\

Let is denote by $D_n$ the space spanned by the $z_n$'s. It can be
viewed as ${\cal H}_0 \oplus {\cal H}_0 $. To define an evolving
observable $\hat{O}(m)$ corresponding to an operator $\hat{O}$ acting
invariantly on all $D_n$'s, consider the $z_m$ member of an physical
state $\{ |\Phi_n\rangle\}$. Operate on it by $\hat{O}_0$ and evolve back
to $D_0$. This will not in general satisfy the consistency condition. Let
${\cal P}$ be a projection operator which will project any element of
$D_0$ on to one corresponding to a physical one. That is, define a new $z_0$ 
whose components satisfy the consistency condition. Evolve this to any $n > 0$.
This clearly defines a physical state. $\hat{O}(m)$ is defined to produce this
physical state from the starting physical state. By construction, these
operators, defined to act on physical states, produces a new physical
state. Introduce the evolution operators $E(0, n): D_0 \rightarrow D_n$
and $E(n, 0): D_n \rightarrow D_0$ given explicitly by,
\begin{eqnarray}
E(0, n) & := & {\mathbb A}(n -1) {\mathbb A}(n - 2) \cdots {\mathbb A}(0)
\nonumber \\
E(n, 0) & := & {\mathbb B}(1) {\mathbb B}(2) \cdots {\mathbb B}(n)
\end{eqnarray}
Then $\hat{O}(m)$ is expressible as, 
\begin{equation}
[\hat{O}(m)z]_n := E(0, n) \circ {\cal P} \circ E(m, 0)
\hat{O} z_m .
\end{equation}

The projection operator can be constructed easily and is uniquely given
by,
\begin{equation}
{\cal P} := \frac{1}{{\mathbb I} + 4 \beta^2 \hat{H}^2} \left(
\begin{array}{cc} {\mathbb I} & 2 i \beta \hat{H} \\ -2 i \beta \hat{H}
& 4 \beta^2 \hat{H}^2 \end{array} \right) .
\end{equation}

In analogy with the continuum case, one may naturally define a physical inner 
product as the $n = 0$ term of the series in equation 8. We can make other
possible choices, for example, $z_0^{\dagger} {\cal M} z_0$, where
${\cal M}$ is a suitable $2 \times 2$ matrix of operators on ${\cal
H}_0$. ${\cal M} = diag( {\mathbb I}, {\mathbb O} )$ will produce the $n
= 0$ term of the series. Physical matrix elements of the evolving observables 
are then given by,
\begin{equation}
\langle\Psi^{\prime}|\hat{O}\Psi\rangle = {z^{\prime}_0}^{\dagger} {\cal P M P} E(m, 0) \hat{O} E(0, m) z_0 
\end{equation}
We have used $z_0^{\dagger} {\cal P} = z_0^{\dagger}$ for physical
states to obtain a symmetrical expression. While this looks very similar
to the continuum case (apart from the presence of ${\cal P M P}$), its
implications are very different.\\

These definitions, while plausible, are unsatisfactory. We have focussed on the
operators which act invariantly on the space of physical states. However
to be of use for measurements, such operators must satisfy further
properties such as self-adjointness. This of course needs to be defined
relative to the physical inner product. The non-unitarity of evolution
however implies that even if have a self-adjoint operator at $n = 0$,
the other members of the family are not self-adjoint in general. The
presence of the projection operator implies that algebraic relations
among operators are not preserved by the evolution. For example, an
operator associated with an $\hat{O}^2(0)$ is {\it not} the square of
the corresponding operator associated with $\hat{O}(0)$ and like wise
for commutation relations. In the present case of Fock representation, the 
condition arises due to the spectrum of number operator being bounded below. 
In LQC, although the state label $n$ takes all integral values, there 
is still the consistency condition which should cause similar difficulties. \\

One could have dealt with the second order equation it self. Now there
is no need for any explicit projection operator. The evolution is still
non-unitary but in addition, one can not evolve a given $|\Phi_n\rangle$ back 
to a $|\Phi_0\rangle$ since the equation is second order. The first order 
formulation avoids this but introduces explicit projection operator. Thus 
our attempt to mimic the steps followed in the continuous time case do not 
lead to satisfactory definitions.\\

However, we can appeal to the relations between the continuum and the 
discrete description discussed before. Then the physical inner product
and matrix elements as defined in the continuous case can be expressed 
as,
\begin{eqnarray}
\langle\Phi^{\prime}(\tau) | \Phi(\tau) \rangle & = & \sum_{m,n}
f_m^*(\tau) f_n(\tau) \langle \Phi^{\prime}_m | \Phi_n \rangle \nonumber \\
\langle\Phi^{\prime}(\tau) | A |\Phi(\tau) \rangle & = & \sum_{m,n}
f_m^*(\tau) f_n(\tau) \langle \Phi^{\prime}_m | A | \Phi_n \rangle 
\end{eqnarray}

These are {\it very} different from the physical inner products and matrix
elements  we attempted previously! The right hand sides involve infinite
sums and are {\it highly non-local} in the discrete time label. Note
that the apparent $\tau$ dependence on the right hand sides is
consistent with that implied by the left hand sides of the above
equations. Thus, {\it if we somehow 
invented these inner products and matrix elements, we could construct a 
continuum description from the discrete one}. In the case of quantum
mechanical example we are discussing, we have the advantage of knowing a
continuum description ab initio but for LQC also something similar can
be conceivable. This however is not attempted in the present work. In
the next section we make some general remarks regarding viability of a
`dynamical' interpretation and attempt to arrive at an interpretation of
$\beta$.

\section{`Evolution' in Quantum Mechanics}

Unlike classical mechanics, quantum mechanics permits {\it two} notions
of evolution which are {\it not} equivalent due to the uncertainty
relation. The two notions correspond to evolution at the level of states
i.e. a continuous or a discrete family of rays (or vectors) and evolution at the
level of observed quantities i.e. family of expectation values of observables.
To appreciate the non-equivalence of these two let us quickly recall the
derivation of time-energy uncertainty relation. One can do this quite
generally. \\

Let $G$ be a self adjoint operator on a Hilbert space. Consider the one
parameter group of unitary operators generated by $G$, $U(\xi) := $
exp($-i \xi G$), $\xi \in {\mathbb{R}}$. Define a family of normalized 
state vectors $|\psi(\xi)\rangle := U(\xi)|\psi(0)\rangle$. For any self
adjoint operator, $A$, corresponding to an observable define
$f_{\psi}(\xi) := \langle \psi(\xi)| A |\psi(\xi)\rangle$. Assume $A$ to
be independent of $\xi$ for simplicity. Then it follows that,
\begin{equation}
\delta f_{\psi}(\xi) := \delta\xi \frac{\partial f_{\psi}}{\partial \xi}
 =  i \delta\xi \langle\psi(\xi)| [ G , A ] |\psi(\xi)\rangle~, ~~~~~~~~~
 \delta\xi ~>~ 0 ~~\mbox{(say)}
\end{equation}

Defining $G^{\prime} := G - \langle\psi(\xi)| G | \psi(\xi)\rangle$ and
like wise for $A$ one gets,
\begin{eqnarray}
|\delta f_{\psi}(\xi) | 
& = & | ~ \delta\xi \langle\psi(\xi)| [ G^{\prime} , A^{\prime} ] |\psi(\xi)\rangle ~ | \nonumber \\
& \le & 2 \delta\xi ~ | \mbox{Im}(G^{\prime}\psi, A^{\prime}\psi) | \nonumber \\
& \le & 2  \delta\xi ~ | (G^{\prime}\psi, A^{\prime}\psi) | \nonumber \\
& \le & 2  \delta\xi ~ || G^{\prime}|\psi\rangle ||~|| A^{\prime}|\psi\rangle || \nonumber \\
& \le & 2 \delta\xi ~ \Delta G_{\psi} ~ \Delta A_{\psi} 
\end{eqnarray}

Clearly in order to detect a change in the expectation value
$f_{\psi}(\xi)$, the change computed above must be at least as large as
the uncertainty, $\Delta A_{\psi}$. This immediately gives the uncertainty
relation:
\begin{equation}
\delta\xi ~ \Delta G_{\psi} \ge \frac{1}{2} 
\end{equation}

Note that this derivation is independent of canonical commutation relations 
and thus is {\it not} tied to a phase space $\sim {\mathbb{R}}^{2N}$, 
$~\delta\xi \ne \Delta \hat{\xi}_{\psi}$ although it could be. The above 
derivation has also assumed $\delta\xi$ to be small enough so that
the higher order terms can be neglected. If these are also included, then
the uncertainty relation will assume a different form. \\

Taking $G$ to be the Hamiltonian  and $\xi = \tau/\hbar$ we get the usual 
time-energy uncertainty relation while for $G$ equal to the momentum
(say) and $\xi = q/\hbar$ we get the position-momentum uncertainty relation. 
Its meaning is that we can not {\it observationally} resolve values between 
$\xi$ and $\xi + \delta\xi$. The fact that there is a non-zero lower
bound and $\Delta G_{\psi}$, in a physical situation is always {\it finite}
(though it could be very large) implies that continuum values for $\xi$
are strictly mathematical idealizations. This distinguishes the two
notions of `evolution' in quantum mechanics, mentioned above. The states
can be thought of as evolving continuously but observationally,
continuous evolution is necessarily an idealization. The
absence of a non-zero lower bound in classical mechanics permits continuum 
values of $\xi$ to be taken more literally. Note that this applies not
just to `time' but also to `space'.\\

The notion of observationally detectable evolution can be articulated as
follows. We can meaningfully say that a system has changed its state
provided we can measure at least one of its properties and detect a
change. Any such measurements will give expectation values together with
uncertainties for the corresponding observable. Thus to conclude that
a system in some given state has changed `over a period of time' one
must be able to find at least one observable whose expectation value
in that state changes more that the uncertainty. Note that this must
be understood at the level of an ensemble of identically prepared
systems since a single measurement on a single system will just produce
some eigenvalue of the observable according to the standard
interpretation of quantum mechanics. To account for `over a period of time', 
one must assume a family (discrete or continuous) of states in which the
expectation values are to computed.\\

Thus, for the observational notion of an evolution, the central quantities are
expectation values. Given a discrete family of vectors and self adjoint operators
(or a family thereof) one can construct a corresponding family of
expectation values. Such a family could be usefully interpreted as an
`evolution' provided that the difference between consecutive members of the 
family of expectation values is larger than the corresponding uncertainties, 
at least for some observable and for generic states. Such a criterion of
{\it detectable} evolution is independent of how the family of vectors is chosen
and whether the members of this family are connected by unitary operators. 
%Unitarity of evolution may be a convenience but does not appear to be a 
%necessity. 
This is also 
independent of whether the sequence of vectors is obtained from solutions of 
some (Hamiltonian) constraint of a constrained system. \\

As seen in the example of quantum mechanical system, the net result of
imposing constraint in the kinematical arena is to produce a family of
vectors in ${\cal {H}}_0$. Since such families are uniquely
determined by $|\Phi_0\rangle$ (or $|\Phi(0)\rangle$), the space of
`physical states' is isomorphic to ${\cal{H}}_0$. A natural choice of
physical inner product is then simply the inner product in ${\cal{H}}_0$ and
`physical observables' are naturally self adjoint operators on
this Hilbert space. We can now construct a sequence of `physical' expectation
values:
\begin{equation}
\langle A \rangle_n := \frac{\langle\Phi_n | A | \Phi_n\rangle}
{\langle\Phi_n | \Phi_n\rangle}
\end{equation}

Expanding the states in terms of eigenstates of the Hamiltonian
(assuming discrete spectrum for simplicity), $|\Phi_n\rangle =
\sum_{\rho} C_{n,\rho}|{\cal{E}}_{\rho}\rangle$, we get,
\begin{equation}
\langle A \rangle_n = \frac{\sum_{\rho, \sigma} C^*_{n,\rho} C_{n,\sigma} 
\langle{\cal{E}}_{\rho} | A | {\cal{E}}_{\sigma}\rangle}
{\sum_{\rho} |C_{n,\rho}|^{2}}
\end{equation}

If $|\Phi_0\rangle$ is an eigenstate of the Hamiltonian, then so are
$|\Phi_n\rangle ~ \forall ~ n$. The expectation values defined above are then
independent of $n$. Thus eigenstates of the Hamiltonian are `stationary'
states even with respect to the discrete `evolution'. \\

The coefficients $C_{n,\rho}$ above, satisfy the difference equations
and are $n^{th}$ order polynomials in $(-2 i \beta {\cal{E}}_{\rho})$.
The expectation values and the uncertainties are thus rational functions
of $\beta$. \\ 

Applying the reasoning to the simplest two level system brings out
further possibilities. Let ${\cal H}_0$ be two dimensional and let the
Hamiltonian be $\hat{H} = {\cal E}\sigma_3$. Let a generic observable be a
Hermitian $2\times2$ matrix. Then it is easy to see that all expectation
values (and hence also the uncertainties) are {\it independent} of
$\beta$! Explicitly, let
\begin{equation}
|\Phi_0\rangle ~:= ~ \left(\begin{array}{l} cos \theta \\ sin 
\theta e^{i\phi}\end{array}\right) ~~~~~~~~~~~
A ~ := ~ \left( \begin{array}{ll} a & b \\ b^* & c \end{array} \right)
~~~~~ \mbox{a, c real.}
\end{equation}

Putting $|\Phi_n\rangle := \frac{P_n}{\sqrt{n!}}|\Phi_0\rangle, ~
\forall n \ge 0$ and $ Q := - 2 i \beta \hat{H}$, one can see that the
$P_n$'s satisfy the equation,
\begin{equation}
P_{n + 2} = Q P_{n + 1} + (n + 1) P_n, ~~~~ \forall n \ge -1, P_0 :=
{\mathbb I}, P_{-1} := 0
\end{equation}

By inspection, $P_n$'s are $n^{th}$ order polynomials in $Q$.
Furthermore, for even (odd) $n$, only even (odd) powers of $Q$ occur. 
Since $Q^2$ is a multiple of the identity matrix, $P_{2m}$ is a
polynomial in $(-4 \beta^2 {\cal{E}}^2)$ times the identity matrix while 
$P_{2m + 1}$ is another polynomial in the same variable times
$\sigma_3$.  In computing the expectation values, these polynomials
cancel out in the ratios and all the $\beta$ dependence disappears
leaving us with,
\begin{eqnarray}
\langle A \rangle_n & = & a ~ cos^2 \theta + c ~ sin^2 \theta + (-1)^n sin
\theta ~ cos \theta ~( b e^{i\phi} + b^* e^{-i \phi} ) \nonumber \\
(\Delta A)^2_n & = & |b|^2 + sin^2 \theta ~ cos^2 \theta ~ \{ (a -c)^2 - 
( b e^{i\phi} + b^* e^{-i \phi} )^2 \} \nonumber \\
& & + ~ (-1)^n sin \theta ~ cos \theta ~ cos 2\theta ~ (c - a)( b e^{i\phi} + b^* e^{-i \phi} )
\end{eqnarray}

From these it follows that change in the consecutive expectation values
depends only on $b$ as it should since the diagonal part of $A$ commutes
with the Hamiltonian. Taking $a = c = 0$ for simplicity and requiring that 
the change be at least as large as the uncertainty leads to,
\begin{equation}
\frac{|~sin \theta ~ cos \theta ( b e^{i\phi} + b^* e^{-i \phi} ) ~|}
{|b|} ~ \ge \frac{1}{\sqrt{5}}
\end{equation}

This simple example illustrates that a discrete evolution of the type
being considered could be independent of $\beta$, could have an
oscillatory n-dependence and there could be a sub-class of states which 
are {\it not} eigenstates of the Hamiltonian and yet will not exhibit
detectable evolution. These are of course special properties of the
particular system.\\

More generically, one could try to see the $\beta$ dependence in the limit
$\beta \rightarrow 0$. The expectation values can then be expressed as a
power series in $\beta$. It turns out that,
\begin{eqnarray}
\langle A \rangle_{2m} & \approx & \langle A \rangle_0 + o(\beta^2)
\nonumber \\
\langle A \rangle_{2m + 1} & \approx & \frac{\langle HAH
\rangle_0}{\langle H^2 \rangle_0} + o(\beta^2)
\nonumber \\
(\Delta A)^2_{2m} & \approx & (\Delta A)^2_0 + o(\beta^2)
\end{eqnarray}

Thus the difference of expectation values at $n = 2m + 1$ and $n = 2m
-1$ is of order $\beta^2$ while the uncertainty at $n = 2m$ is the
uncertainty at $n = 0$ plus a term of order $\beta^2$. For detectability then 
the uncertainty at $n = 0$ must be comparable to $\beta^2$. This gives a
hint about the role of $\beta$. If we select a set of observables with
respect to which we wish to detect an evolution then $\beta$ should be
chosen to be of the order of (or larger than) the square root of the 
uncertainty of the observable with the smallest uncertainty. Note that
this gives a {\it lower limit} on the value of $\beta$. In this manner, 
the criterion of detectable evolution can be used to get some condition(s) 
on $\beta$. However there does not seem to be a simple way of obtaining an 
analogue of uncertainty relation.  For continuous family $|\Phi(\tau)\rangle$,
there is no $\beta$, the evolution is unitary and the usual results follow.\\

{\underline {Remarks:}}\\

1) The possibility of a discrete time arises naturally for a theory
presented in a frozen time form. Even conventionally presented theories
can be cast in this form and we exploited this in constructing our
example. For such theories (always a constrained theory), one needs to choose 
a suitable degree of freedom as a `time degree of freedom'. One can
always view the kinematical Hilbert space as a tensor product of Hilbert
space of the time degree of freedom and the Hilbert space of the rest of
the degrees of freedom. Solutions of the constraint can then be obtained
as a families of vectors in the non-time sector. The `time' now appears as
a label for each of the family and this could be continuous or discrete. The
families themselves are then determined as  solutions of differential or
difference equation. The form and order of the equations depends on the
form of the constraint, the choice of representation (or choice of basis
in the Hilbert space of the time degree of freedom) and of course on the
choice of the time degree of freedom. Except for these details, a
discrete time presentation can be set up generally. \\

A continuum {\it approximation} for a discrete presentation can be looked for
in the usual manner as indicated in subsection II B. The stronger notion
of `pre-classicality' however may not always be realizable. Even in our
case, we do have {\it asymptotic} solutions which do have a pre-classical
limit but {\it exact} solutions do not admit such a limit. Emergence of
continuous time however can be sought via a transform instead of a
limit.\\

Note that we did not need to be explicit about the Hilbert space of non-time 
degrees of freedom.  Even the number of degrees of freedom is unimportant 
for discrete time description. These details are of course crucial in the 
construction of ${\cal H}_0$ and observables. \\

2) The parameter $\beta$ is a priori completely arbitrary and has dimensions 
of inverse energy, $\hbar\beta$ is thus a time scale. What possible 
interpretation can one ascribe to this parameter? In particular, does it 
reflect some intrinsic property of the system (and thus is selected by the 
system) or is it related to the resolutions with which a set of measurements 
are performed (and thus is to be selected by the experimenter)? Quite
independently, there are two time scales: the intrinsic one by which a physical 
system keeps evolving and the clock scale eg. the least count of an actual 
clock of an observer. \\

If $\hbar\beta$ is the intrinsic time scale then logic of continuum
approximation would require that the clock scale be much much larger that
the intrinsic one so that continuum time description is a very good
approximation. Conversely if the clock scale is comparable to the
intrinsic scale then one should use the discrete evolution. The schematic
argument for small $\beta$ given above would now imply that discrete
evolution may still not be observable if the uncertainties in the
tracked observables are larger than permitted by the $\beta$.\\

The intrinsic time scale may be roughly estimated to
be of the order of the inverse of the {\it maximum} uncertainty in
energy measurement that may occur in the system. This need not be infinite, 
since there would be a maximum energy above which modeling of the physical 
system breaks down - eg. particle a box would not be a valid description for
arbitrarily high energies though the spectrum of the Hamiltonian is of
course unbounded. \\

If however $\hbar\beta$ is not an intrinsic time scale, then it needs to
be adjusted depending upon what observables are used for tracking
evolution. The small $\beta$ argument gives lower bounds on $\beta$ in
terms of the uncertainties of the tracked observables. Unlike the
continuous evolution, which is also unitary, the detectability of
evolution is directly dependent on particular observables used for
tracking. \\

We are unable to decide between the alternatives. It is possible that
$\hbar\beta$ is a scale intermediate to the intrinsic and the clock
scales. Considering analogy with the Immirzi parameter in the context of
LQC, interpretation of $\beta$ here may also throw some hints about the
role of the Immirzi parameter.\\

3) One of the powerful methods of studying semi-classical limit for
systems with phase space $~ R^{2N}$, is via the Wigner distribution
function \cite{berry}. It is essentially a double Fourier transform of 
expectation value of a certain unitary operator. In the usual continuous time
description, the distribution function satisfies the classical Liouville
equation to leading order in $\hbar$ when the expectation value is taken
with respect to a solution of the Schrodinger equation \cite{halliwell}.
It would be interesting to repeat the steps with discrete time though it
looks complicated because of the structure of the difference equation. \\

4) This work has been motivated by the LQC work. So what does it say
about discrete time evolution in LQC? As mentioned in the introduction,
at present the discussion of physical quantities such as inner products
and matrix elements of observables is at a schematic level. The present
work points out the possible pitfalls one may encounter. Our analysis of
pre-classical limit indicates that such a limit could exist, at the level
of states, only for solutions whose asymptotic behaviour has a
characteristic root equal to 1. It may still go through at the level of
expectation values if the largest root is positive. For the isotropic
cosmology with positive spatial curvature (Bianchi-IX), for the
expressions given in \cite{isotropic}, there is neither a root equal to one 
nor is the highest root positive. However, independent of whether a 
pre-classical limit exist or not, one could go ahead with a discrete equation 
at the level of expectation values. One may then take recourse to Wigner
distribution formalism to explore the semi-classical limit. This of
course needs the Wigner distribution formalism to be developed in the
context of the polymer representation. \\

While one may construct families of states and even arrange schemes to
distinguish corresponding expectation values, it leaves unanswered the
question as to why does any system evolve at all?

\begin{acknowledgments}
I would like to thank Martin Bojowald for useful comments and pointing
out the earlier works on discrete time quantum mechanics.
\end{acknowledgments}


\begin{thebibliography}{99}

\bibitem{wd-oprs}
M. Bojowald 2001, {\it Class. Quant. Grav.} {\bf 18}, 1055; 
gr-qc/0008052.

\bibitem{discrete}
M. Bojowald 2001, {\it Class. Quant. Grav.} {\bf 18}, 1071;
gr-qc/0008053.

\bibitem{isotropic} M. Bojowald 2002, Class. Quant. Grav. {\bf
19}, 2717; gr-qc/0202077.

\bibitem{absense}
M. Bojowald 2001, {\it Phys. Rev. Lett.} {\bf 86}, 5227;
gr-qc/0102069.

\bibitem{unique}
M. Bojowald 2001, {\it Phys. Rev. Lett.} {\bf 87}, 121301;
gr-qc/0104072.

\bibitem{semiclassical}
M. Bojowald 2001, {\it Class. Quant. Grav.} {\bf 18}, L109;
gr-qc/0105113.

\bibitem{ambiguities} M. Bojowald 2002, gr-qc/0206053.

\bibitem{history}
T. D. Lee 1983, {\it Phys. Lett} {\bf 122B}, 217; \\
R. Friedberg and T. D. Lee 1983, {\it Nucl. Phys.} {\bf B225 [FS9]},
1;\\
C. M. Bender, K. A. Milton, D. H. Sharp, L. M. Simmons,Jr and R. Stong
1985, {\it Phys. Rev} {\bf D 32}, 1476; \\
C. M. Bender, L. R. Mead and K. A. Milton 1993,  hep-ph/9305246; \\
G. Jaroszkiewicz and K. Norton 1997, {\it J. Phys.} {\bf A 30}, 3115, hep-th/9703079;\\
G. Jaroszkiewicz and K. Norton 1997, {\it J. Phys.} {\bf A 30}, 3145,
hep-th/9703080;\\
G. Jaroszkiewicz and K. Norton 1998, {\it J. Phys.} {\bf A 31}, 977, hep-th/9707029;\\
G. Jaroszkiewicz and K. Norton 1998, {\it J. Phys.} {\bf A 31}, 1001,
hep-th/9707030;\\
G. Jaroszkiewicz and K. Norton 1998, hep-th/9804165;\\
C. Di Bartolo, R. Gamibini and J. Pullin 2002, gr-qc/0205123.

\bibitem{evolving}
A. Ashtekar, R.S. Tate and C. Uggla 1993, {\it Int. Jour. Mod. Phys.}
{\bf D2}, 15; gr-qc/9302027.

\bibitem{diff-eqns} Saber N. Elyadi 1996, {\it An Introduction to Difference
Equations} (Springer-Verlag, New York).

\bibitem{berry} M. V. Berry 1977, {\it Phil. Trans. Roy. Soc.} {\bf A 287},
237.

\bibitem{halliwell} J. J. Halliwell 1987, {\it Phys. Rev.} {\bf D 36},
3626.

\end{thebibliography}
\end{document}